\documentclass[english]{article}
\usepackage[T1]{fontenc}
\usepackage[latin1]{inputenc}
\usepackage{amssymb}

\makeatletter

\usepackage{float}
\usepackage{graphicx}

\usepackage{babel}
\makeatother
\begin{document}
\renewcommand{\thefootnote}{$*$}
\noindent
{\Huge \sf Casimir Effect in Compact Universes}
\vspace{1cm} \\
{\large Daniel M\"uller}\footnote{muller@fis.unb.br} \\
Instituto de Física - Universidade de Brasília \\ 
Campus Universitário Darcy Ribeiro\\
Cxp 04455, 70919-900, Brasília, DF, Brasil
\section*{Abstract}
The Casimir effect in compact hyperbolic Universes was numerically obtained 
in previous publications. In this talk, I expose these results.

\section{Introduction}

The Universe today is very close to homogeneous. Also, no preferred
direction has been discovered, so that the most important 3-Manifolds
in connection with cosmology are $S^{3},\: H^{3}$ and $R^{3}.$ Anyway,
it must be mentioned the following interesting results, that isotropisation can be 
achieved by particle production in the primordial Universe 
\cite{lukash1974,zeldovich1972} or by
the existence of a classical cosmological constant \cite{wald1983}.
For the moment, let us simply suppose that the Universe really is
homogeneous and isotropic, which is in good agreement with the observations.
It is well known that $R\times S^{3},\: R\times H^{3}$ and $R\times R^{3}$
are solutions of Einstein's equations (EQ). It is irrelevant for the EQ, if the 
spatial sections have a non trivial topology modelled on them, or not. From recent determinations
of the WMAP mission there is an indication for a spherical Universe
\cite{bennett2003}, but still not capable of ruling out the other
two cases. The low multipoles of the CMBR, are the ones more sensible to 
global properties of the space. In a very interesting result, \cite{luminet2003}
showed that the low multipoles of the CMBR are better fitted, if the
spatial sections of the Universe are the Poincaré dodecahedral space,
instead of $S^{3}$ (see also \cite{ellis2003}).

There are many effects that arise if space is compact. The first astrophysical
limits on the topology of the universe were obtained for a 3-torus
$T^{3}$. Accordance with the homogeneity of the CMBR puts a lower
limit on the size of the fundamental cell, about $3000$ Mpc, which
is a cube in the cases of \cite{sok} and \cite{as}. Later on, it
was shown that this result is very sensitive to the type of the compactifications
of the spatial sections. For a universe with spatial sections $T^{2}\times R$,
the fundamental cell's size is about $1/10$ of the horizon, and is
compatible with the homogeneity of the cosmic microwave background
radiation CMBR \cite{rouk}. In compact universes, the pair separation
histogram would present spikes for characteristic distances. At first
it was thought that this technique, known as the \textit{crystallographic
method}, was able to totally determine the topology of the universe
\cite{lll}. It turned out that the crystallographic method only applied
when the holonomy group contained at least a Clifford translation,
i.e. a translation which moves all the points by the same distance
\cite{llu} and \cite{gtrb}. Generalizations of the crystallographic
method were proposed, for example in \cite{fg}. 

Also in compact universes the light front of the CMBR interacts with
itself producing circles in its sky pattern \cite{css2}. 

Recent results have called our attention to the possibility that methods
based on multiple images will prove not to be efficient \cite{grt}.
The reasoning is that, according to observations, the curvature is
very small, so the fundamental regions are so big that there has not
been time enough for the formation of ghost images. The result is
that for low curvature universes such as ours, only compact universes
with the smallest volumes could be detected by pattern repetitions. 

The knowledge of the spectrum of the Laplacian for compact spaces
have some motivations. One of them is the decomposition of the CMBR
in right set of modes, if space is non trivial. It has been studied
numerically for compact hyperbolic space by \cite{cornish1998,inoue1998}.
And more recently \cite{lehoucq2002} for compact spherical space. 

Casimir effect also occurs in compact spaces. In the most simple cases
it can be obtained by analytic continuation of the zeta function.
This procedure is known as the zeta function regularization, see for
example \cite{bordag2001}, \cite{elizalde.et.al.}. Very elegant
analytical results can be obtained for the Casimir effect in hyperbolic
space \cite{goncharov1990,bytsenko1991}. The method is indirect,
as it makes use of the Selberg trace formula to obtain the zeta function,
irrespective of the knowledge of the spectrum. For a more comprehensive
introduction to the Selberg trace formula see \cite{balazsvoros}.
There is a recent result for the Casimir energy in shperical space
\cite{elizalde2004}.

In this work I shall expose a different technique which was previously
used by us \cite{muller2001,muller2002a,muller2002b}. 

The paper is organized as follows. In section \ref{ectop}, I review
the simplest manifestation of the Casimir effect of topological nature,
and reproduce some basic properties of zeta functions. The static
solution of EQ, the classical energy momentum tensor
and the method of point splitting is outlined in section \ref{s1}.
In section \ref{propf} the Feynman propagator and the Hadamard function
is obtained. And in section \ref{cefu} the Casimir energy density
is numerically obtained in some Universes. Natural units are used, 
$\hbar=G=c=1$, except in section \ref{s1}. 

\section{Topological Casimir Effect\label{ectop}}

It was discovered by Casimir in 1948 \cite{casimir1948} that two
uncharged parallel metallic plates in vacuum should be attracted by
the force \[
f=-\hbar c\frac{\pi^{2}}{240}\frac{S}{a^{4}},\]
 where $S$ is the area of the plates and $a$ is the distance between
them. Soon after, it was verified experimentally \cite{sparnay1958}.
It is universal since it does not contain the coupling constant of
the field. Also, it is a non local effect, in the sense that it depends
on the boundary conditions.

Casimir effect also occurs when the topology of space is non trivial.
In the following, let us reproduce the result for a massless, uncharged
scalar field in the two-dimensional space $R\times S^{1}$, for more
details see \cite{grib1994,elizalde.et.al.,bordag2001}. In this case
the spectrum of the Klein-Gordon operator reads \[
\omega_{n}^{2}-k_{n}^{2}=0,\:\: k_{n}=\frac{2\pi}{a}n,\:\: n=0,\pm1,\pm2,...\]
 where $a$ is the \textit{volume} of the manifold. The total energy
is \[
\left\langle b\left|H\right|b\right\rangle =\sum_{k}\omega_{k}\left(n_{k}+\frac{1}{2}\right),\]
where $\left|b\right\rangle $ is an arbitrary state of the field, 
with $n_k$ particles in mode $k$. The
vacuum expectation value follows from considering $n_{k}=0,$ and
the Casimir energy reads \begin{eqnarray*}
\left\langle 0\left|H\right|0\right\rangle  & = & \frac{1}{2}\sum_{k}\omega_{k}\\
E_{c} & = & \frac{2\pi}{a}\sum_{n=1}^{\infty}n.\end{eqnarray*}
 As it stands, the last summation of course diverges. Nevertheless,
a meaning can be given to it through the mathematically well known
Riemann zeta function. Remind that the Riemann zeta function is defined
for $Re(s)>1$ \[
\zeta(s)=\sum_{n=1}^{\infty}\frac{1}{n^{s}}.\]
This function can be analytically continued giving a finite result
\cite{whittaker1963} \[
\zeta(-1)=-\frac{1}{12}.\]
The Casimir energy for $R\times S^{1},$ then follows \[
E_{C}=-\frac{\pi}{6a}.\]
The technique used above is the famous zeta function renormalization. 
Apparently, Zeta functions where used in the physics literature for the first 
time in \cite{dowker1976,hawking1977}.
In the following I shall expose the well known relation between the
zeta function and the trace of the heat kernel. Suppose a Laplace
type operator A acting on a compact 3-space, with a spectrum 
$\lambda_{1},\:\lambda_{2},$$\:...$\[
A\phi_{n}=\lambda_{n}\phi_{n},\]
where $\phi_n$ is some complete orthonormal set of modes.
The Euler infinite integral for the gamma function is defined as

\[
\Gamma(s)=\int_{0}^{\infty}dxx^{s-1}e^{-x}.\]
Now consider the change of variables $x=\lambda_{n}t$ \[
\Gamma(s)=\lambda_{n}^{s}\int_{0}^{\infty}dtt^{s-1}e^{-t\lambda_{n}},\]
which can be inverted to give $\zeta_{A}(s),$ the zeta function of
the 3-manifold\begin{eqnarray*}
\sum_{n}\frac{1}{\lambda_{n}^{s}} & = & \frac{1}{\Gamma(s)}\int_{0}^{\infty}dtt^{s-1}\sum_{n}e^{-t\lambda_{n}}\\
\zeta_{A}(s) & = & \frac{1}{\Gamma(s)}\int_{0}^{\infty}dtt^{s-1}Y_{A}(t).\end{eqnarray*}
In this last formula, \[
Y_{A}(t)=\sum_{n}e^{-t\lambda_{n}},\]
is the trace of the heat kernel $K(t,x,x^{\prime})$,
\begin{eqnarray*}
& &-\frac{\partial}{\partial t}K(t,x,x^{\prime}) = A_{x}K(t,x,x^{\prime})\\
& &K(t,x,x^{\prime}) = \sum_{n}e^{-t\lambda_{n}}\phi_{n}(x)\phi_{n}(x^{\prime})\\
& &Y_{A}(t) = \int_{V_{\mathcal{M}}}d^{3}x\sqrt{g}K(t,x,x).
\end{eqnarray*}
Where the subscript $x$ means that the operator $A$ do not act on
the variable $x^{\prime}$, $g$ and $V_{\mathcal{M}}$  is the determinant of the
metric and  the volume of the 3-manifold, respectively. The parameter $t$ 
above is not to be confused with time.

\section{Quantum Field Theory in The Static Universe $R\times H^{3}$ \label{s1}}

The hyperbolic space sections $H^{3}$, can be realized as a surface
\begin{equation}
\left(x-x^{\prime}\right)^{2}+\left(y-y^{\prime}\right)^{2}+\left(z-z^{\prime}\right)^{2}-\left(w-w^{\prime}\right)^{2}=-a^{2},\label{cond}\end{equation}
 embedded in a Minkowski 4-space \[
dl^{2}=dx^{2}+dy^{2}+dz^{2}-dw^{2}.\]
 As this space is homogeneous, we explicitly write the origin of coordinates
$(x^{\prime},y^{\prime},z^{\prime},w^{\prime})$. It can easily be
seen that its isometry group is the proper, orthocronous Lorentz group
$SO^{\uparrow}(1,3),$ which is isomorphic to $PSL(2,C)=SL(2,C)/\{\pm I\}$.
With the constraint of Eq. (\ref{cond}) on the line element we obtain
\begin{eqnarray}
dl^{2} & = & dx^{2}+dy^{2}+dz^{2}\nonumber \\
 &  & -\frac{\left(\left(x-x^{\prime}\right)dx+\left(y-y^{\prime}\right)dy+\left(z-z^{\prime}\right)dz\right)^{2}}{\left(x-x^{\prime}\right)^{2}+\left(y-y^{\prime}\right)^{2}+\left(z-z^{\prime}\right)^{2}+a^{2}},\nonumber \\
ds^{2} & = & -dt^{2}+dl^{2}=g(x,x^{\prime})_{\mu\nu}dx^{\mu}dx^{\nu},\label{elxp}\end{eqnarray}
 where we interchangeably write $\left(x^{0},x^{1},x^{2},x^{2}\right)\longleftrightarrow\left(t,x,y,z\right)$.
Both connections, $\nabla_{x}$ and $\nabla_{x^{\prime}}$, compatible
with the metric of Eq. (\ref{elxp}), can be defined through \ \begin{eqnarray}
 &  & \nabla_{\mu}g(x,x^{\prime})_{\alpha\beta}\equiv0\,\,,\label{cn1}\\
 &  & \nabla_{\mu^{\prime}}g(x,x^{\prime})_{\alpha\beta}\equiv0\,\,.\label{cn2}\end{eqnarray}
 The expression in Eq. (\ref{elxp}) is the popular Robertson-Walker
line element, which written in the Lobatchevsky form reads \begin{equation}
ds^{2}=-dt^{2}+a^{2}\left[d\chi^{2}+\sinh^{2}\chi\left(d\theta^{2}+\sin^{2}\theta d\phi^{2}\right)\right]\,\,,\label{ell}\end{equation}
 with \  \[
\sinh^{2}\chi=\frac{\left(x-x^{\prime}\right)^{2}+\left(y-y^{\prime}\right)^{2}+\left(z-z^{\prime}\right)^{2}}{a^{2}}\,\,.\]

As is well known, the EQ for the homogeneous and isotropic space sections
in Eq. (\ref{ell}), with $a=a(t)$, reduce to the Friedmann-Lema\^{\i}tre
equations \begin{eqnarray*}
 &  & \left(\frac{\dot{a}}{a}\right)^{2}-\frac{1}{a^{2}}=\frac{8\pi G}{3}\rho+\frac{\Lambda}{3}\,\,,\\
 &  & 2\left(\frac{\ddot{a}}{a}\right)+\left(\frac{\dot{a}}{a}\right)^{2}-\frac{1}{a^{2}}=-8\pi Gp+\Lambda\,\,,\end{eqnarray*}
 where the right-hand side comes from the classical energy-momentum
source for the geometry, $T^{\mu\nu}=(\rho+p)u^{\mu}u^{\nu}+pg^{\mu\nu}$,
plus the cosmological constant term $\Lambda g^{\mu\nu}$.

We assume that the universe was radiation dominated near the Planck
era, hence $p=\rho/3$, and obtain the following static solution \begin{eqnarray}
 &  & a=\sqrt{\frac{3}{2|\Lambda|}}\,\,,\nonumber \\
 &  & \rho=\frac{\Lambda}{8\pi G}\,\,,\nonumber \\
 &  & ds^{2}=-dt^{2}+a^{2}\left[d\chi^{2}+\sinh^{2}\chi\left(d\theta^{2}+\sin^{2}\theta d\phi^{2}\right)\right]\,\,,\label{elpc}\end{eqnarray}
 where the cosmological constant is negative.

We now wish to evaluate the vacuum expectation value of the energy
density for the case of a universe consisting of a classical radiation
fluid, a cosmological constant, and a non-interacting quantum scalar
field $\phi$. The solution of EQ is given in Eq. (\ref{elpc}), where
the quantum back reaction is disregarded. We use the point splitting
method in the universal covering space $R\times H^{3}$, for which
the propagator is exact. The point splitting method was constructed
to obtain the renormalized (finite) expectation values for the quantum
mechanical operators. It is based on the Schwinger formalism \cite{Schwinger},
and was developed in the context of curved space by DeWitt \cite{dWitt}.
Further details are contained in the articles by Christensen \cite{chris1},
\cite{chris2}. For a review, see \cite{grib1994}.

Metric variations in the scalar action \[
S=-\frac{1}{2}\int\sqrt{-g}(\phi_{,\rho}\phi^{,\rho}+\xi R\phi^{2}+m^{2}\phi^{2})d^{4}x\,\,,\]
 with conformal coupling $\xi=1/6,$ give the classical energy-momentum
tensor \begin{eqnarray}
T_{\mu\nu} & = & \frac{2}{3}\phi_{,\mu}\phi_{,\nu}-\frac{1}{6}\phi_{,\rho}\phi^{,\rho}g_{\mu\nu}-\frac{1}{3}\phi\phi_{;\mu\nu}\nonumber \\
 &  & +\frac{1}{3}g_{\mu\nu}\phi\Box\phi+\frac{1}{6}G_{\mu\nu}\phi^{2}-\frac{1}{2}m^{2}g_{\mu\nu}\phi^{2}\,\,,\label{tmunu}\end{eqnarray}
 where $G_{\mu\nu}$ is the Einstein tensor. As expected for massless
fields, it can be verified that the trace of the above tensor is identically
zero if $m=0.$ Variations with respect to $\phi$ result in the curved
space generalization of the Klein-Gordon equation, \begin{equation}
\square\phi-\frac{R}{6}\phi-m^{2}\phi=0\,\,.\label{ekg}\end{equation}

The renormalized energy-momentum tensor involves field products at
the same spacetime point. Thus the idea is to calculate the averaged
products at separate points, $x$ and $x^{\prime}$, taking the limit
$x^{\prime}\rightarrow x$ in the end. \begin{equation}
\langle0|T_{\mu\nu}\left(x\right)|0\rangle=\lim_{x^{\prime}\rightarrow x}T(x,x^{\prime})_{\mu\nu}\,\,,\label{Tnl}\end{equation}
 with

\begin{eqnarray}
 &  & T(x,x^{\prime})_{\mu\nu}\nonumber \\
 & = & \left[\frac{1}{6}\left(\nabla_{\mu}\nabla_{\nu^{\prime}}+\nabla_{\mu^{\prime}}\nabla_{\nu}\right)-\frac{1}{12}g(x)_{\mu\nu}\nabla_{\rho}\nabla^{\rho^{\prime}}\right.\nonumber \\
 &  & \left.\left.-\frac{1}{12}\left(\nabla_{\mu}\nabla_{\nu}+\nabla_{\mu^{\prime}}\nabla_{\nu^{\prime}}\right)+\frac{1}{48}g(x)_{\mu\nu}\left(\square+\square^{\prime}\right)\right.\right.\label{Tmn}\\
 &  & \left.\left.+\frac{1}{12}\left(R(x)_{\mu\nu}-\frac{1}{4}R(x)g(x)_{\mu\nu}\right)-\frac{1}{8}m^{2}g(x)_{\mu\nu}\right]\right.\nonumber \\
 &  & G^{(1)}(x,x^{\prime})\,\,,\nonumber \end{eqnarray}
 where the covariant derivatives are defined in Eqs. (\ref{cn1})
and (\ref{cn2}), and $G^{(1)}$ is the Hadamard function, which is
the expectation value of the anti-commutator of $\phi(x)$ and $\phi(x^{\prime})$
(see below). We stress that the quantity $T(x,x^{\prime})_{\mu\nu}\,\,$only
makes sense after the limit in Eq. (\ref{Tnl}) is taken. The geometric
quantities such as the metric and the curvature are regarded as classical
entities. $g(x)_{\mu\nu}=g(x,x^{\prime}=0)_{\mu\nu}$ is obtained
from the line element in Eq. (\ref{elxp}).

The causal Green function, or Feynman propagator, is obtained as \[
G(x,x^{\prime})=i\langle0|T\phi(x)\phi(x^{\prime})|0\rangle\,\,,\]
 where $T$ is the time-ordering operator. Taking its real and imaginary
parts, \begin{equation}
G(x,x^{\prime})=G_{s}(x,x^{\prime})+\frac{i}{2}G^{(1)}(x,x^{\prime})\,\,,\label{dfg}\end{equation}
 we get for the Hadamard function \[
G^{(1)}(x,x^{\prime})=\langle0|\{\phi(x),\phi(x^{\prime})\}|0\rangle\
=2\mathop{\textrm{Im}}G(x,x^{\prime})\,\,.\]

\section{The Feynman Propagator and the Casimir Energy Density in $R\times\mathcal{M}$\label{propf}}

Green functions, as any other function defined in the spatially compact
spacetime $R\times\mathcal{M}$, must have the same periodicities
of the manifold $\mathcal{M}$ itself. One way of imposing this periodicity
is by determining the spectrum of the Laplacian, which can only be
done numerically.

Another method imposes the periodicity by brute force, \[
f_{\mathcal{M}}(x)=\sum_{\gamma\in\Gamma}f(\gamma x)\,\,.\]
 The above expression is named the Poincar\'{e} series, and when
it converges, it defines functions $f_{\mathcal{M}}$ on the manifold
$\mathcal{M}$.

We define the operator \begin{equation}
F(x,x^{\prime})=F(x)/\sqrt{-g}\delta(x-x^{\prime})\,\,,\label{okf}\end{equation}
 where $F(x)=\square-R/6-m^{2},$ and introduce an auxiliary evolution
parameter $s$ and a complete orthonormal set of states $|x\rangle$,
such that \begin{eqnarray}
 &  & G(x,x^{\prime})=\langle x|\hat{G}|x^{\prime}\rangle\,\,,\nonumber \\
 &  & F(x,x^{\prime})=\langle x|\hat{F}|x^{\prime}\rangle\,\,,\nonumber \\
 &  & \hat{G}=i\int_{0}^{\infty}e^{-is\hat{F}}ds\,\,.\label{iokf}\end{eqnarray}
 This last equation implies that $\hat{G}=(\hat{F}-i0)^{-1}$, hence
the causal Green function becomes \begin{equation}
G(x,x^{\prime})=i\int_{0}^{\infty}ds\langle x|\exp(-is\hat{F})|x^{\prime}\rangle\,\,,\label{int}\end{equation}
 and the matrix element $\langle x|\exp(-is\hat{F})|x^{\prime}\rangle=K(s,x,x^{\prime})$
satisfies a Schr\"{o}dinger type equation, \[
i\frac{\partial}{\partial s}K(s,x,x^{\prime})=\left(\square-\frac{R}{6}-m^{2}\right)K(s,x,x^{\prime})\,\,,\]
where in analogy with section \ref{ectop}, $K$ is named the Schrödinger
kernel.

Assuming that $K$ depends only on the total geodesic distance $-(t-t^{\prime})^{2}+a^{2}\chi^{2}$,
with the spatial part $a^{2}\chi^{2}$ derived from Eq. (\ref{ell}),
the above equation can be solved, and we get \begin{equation}
K(s,x,x^\prime)=\frac{-i\chi}{\sinh\chi}\frac{\exp\{ im^{2}s+i[(t-t^{\prime})^{2}-a^{2}\chi^{2}]/4s\}}{(4\pi\, s)^{2}}.\label{nschroedinger}\end{equation}
 Substituting this solution for the integrand in Eq. (\ref{int}),
gives for the Feynman propagator \begin{equation}
G(x,x^{\prime})=-\frac{m}{8\pi}\frac{\chi}{\sinh\chi}\frac{H_{1}^{(2)}\left(m\sqrt{\left(t-t^{\prime}\right)^{2}-a^{2}\chi^{2}}\right)}{\sqrt{\left(t-t^{\prime}\right)^{2}-a^{2}\chi^{2}}}\,\,,\label{ff}\end{equation}
 where $H_{1}^{(2)}$ is the Hankel function of the second kind and
order one.

The Hadamard function can be obtained from Eqs. (\ref{ff}) and (\ref{dfg}),
\begin{equation}
G^{(1)}(x,x^{\prime})=\frac{m}{2\pi^{2}}\frac{\chi}{\sinh\chi}\frac{K_{1}\left(m\sqrt{-\left(t-t^{\prime}\right)^{2}+a^{2}\chi^{2}}\right)}{\sqrt{-\left(t-t^{\prime}\right)^{2}+a^{2}\chi^{2}}},\label{Hdrmf}\end{equation}
 where $K_{1}$ is the modified Bessel function of the second kind
and order one. The massless limit $m=0$ can immediately be checked:
\[
G^{(1)}(x,x^{\prime})_{m=0}=\frac{\chi}{2\pi^{2}\sinh\,(\chi)}\left\{ \frac{1}{-(t-t^{\prime})^{2}+a^{2}\chi^{2}}\right\} .\]
 Remembering that for $a\rightarrow\infty$ \[
\sinh\chi=a^{-1}\sqrt{(x-x^{\prime})^{2}+(y-y^{\prime})^{2}+(z-z^{\prime})^{2}}\rightarrow\chi\,\,,\]
 the well known Minkowski result is recovered for the massive and
massless cases, \begin{eqnarray*}
 &  & G^{(1)}(x,x^{\prime})=\frac{m}{2\pi^{2}}\frac{K_{1}\left(m\sqrt{-\left(t-t^{\prime}\right)^{2}+r^{2}}\right)}{\sqrt{-\left(t-t^{\prime}\right)^{2}+r^{2}}}\,\,,\\
 &  & G^{(1)}(x,x^{\prime})_{m=0}=\frac{1}{2\pi^{2}}\left\{ \frac{1}{-(t-t^{\prime})^{2}+r^{2}}\right\} ,\end{eqnarray*}
 where $r$ is the geodesic distance in the spatial euclidean section.

Substituting Eq. (\ref{Hdrmf}) and the covariant derivatives (\ref{cn1})
and (\ref{cn2}) into Eq. (\ref{Tmn}), we obtain $T(x,x^{\prime})_{\mu\nu}$.

The Klein-Gordon equation remains unchanged under isometries, \[
\mbox{$\pounds_{\xi}$}\left[\left(\square-\frac{R}{6}-m^{2}\right)\phi\right]=\left(\square-\frac{R}{6}-m^{2}\right)\mbox{$\pounds_{\xi}$}\phi\,\,,\]
 where $\mbox{$\pounds_{\xi}$}$ is the Lie derivative with respect
to the Killing vector $\xi$ that generates the isometry, hence summations
in the Green functions over the discrete elements of the group $\Gamma$
is well defined.

In $\mathcal{M}=H^{3}/\Gamma$, a summation over the infinite number
of geodesics connecting $x$ and $x^{\prime}$ is obtained by the
action of the elements $\gamma\in\Gamma$, which are the generators
$g_{i}$ and their products (except the identity), on $x^{\prime}$.
Since $\Gamma$ is isomorphic to $\pi_{1}(\mathcal{M})$ each geodesic
linking $x$ and $x^{\prime}$ in $\mathcal{M}$ is lifted to a unique
geodesic linking $x$ and $\gamma x^{\prime}$\ in $H^{3}$. Thus
from Eq. (\ref{Tnl}) we get \begin{eqnarray}
\rho_{C} & = & \left\langle 0\left|T(x)_{\mu\nu}\right|0\right\rangle _{\mathcal{M}}u^{\mu}u^{\nu}\nonumber \\
 & = & u^{\mu}u^{\nu}\lim_{x^{\prime}\rightarrow x}\sum_{\gamma\neq1}T\left(x,\gamma x^{\prime}\right)_{\mu\nu}\,\,.\label{resl}\end{eqnarray}
 The infinite summation occurs because the spacetime $R\times\mathcal{M}$
is static, so there has been enough time for the quantum interaction
of the scalar field with the geometry to travel any distance. Since
we know the universe is expanding, the infinite summation is not physically
valid. The presence of the mass term, however, naturally introduces
a cutoff.

In Eq. (\ref{resl}) the subscript $\gamma\neq1$ means that the direct
path is not to be taken into account. It can be shown that this procedure
is equivalent to a renormalization of the comsological constant, see
for instance \cite{muller2002a}.

\section{Casimir Density \protect{\large $\rho_{C}$} in a Few Universes
\label{cefu}}

There are some reasons that turn smaller Universes more interesting.
We describe some spatially compact universes, with increasing volumes,
in subsections \ref{A}-\ref{D}. 

The values of $\rho_{C}$ shown for each manifold were taken at points
$(\theta,\varphi)$ on the surface of a sphere inside its fundamental
region. For all of them the radius of the sphere is the same, $d=a\chi=0.390035...a$,
where $d$ is the geodesic distance. Our result is displayed in Figures
\ref{2}, \ref{3}, \ref{4} and \ref{5} for a scalar field with
mass $m=0.4$, and a metric scale factor $a=10$. Angles $\theta$
and $\varphi$ correspond to the co-latitude and longitude, so the
lines $\theta=0$ and $\theta=\pi$ correspond to the poles of the
chosen sphere. For each manifold we also write explicitly the radius
of the inscribed sphere $R_{inradius}$.

The description that follows applies to all subsections \ref{A}-\ref{D}.
The $g_{i}$ matrices that generate $\Gamma$ were obtained with the
computer program SnapPea \cite{sp}. To yield a numerical result,
the infinite summation (\ref{resl}) has to be truncated. Recall that
this summation occurs in the covering space $H^{3}$. We halted the
summation each time the action of the generators $g_{i}$ and their
products on the origin $(x^{1},x^{2},x^{3})=(0,0,0)$, reached a geodesic
distance bigger than $d=a\chi=5.29834...a$. Care was taken so that
no point was summed more than once. In other words, the summation
(\ref{resl}), which yelds $\rho_{C}$, was truncated when the interior
of the hyperbolic sphere of geodesic radius $d=a\chi=5.29834...a$,
was covered with replicas of the fundamental region. We checked that
additional contributions were unimportant for the evaluation of $\rho_{C}$.

\subsection{Weeks Manifold \label{A}}

\begin{figure}[H]
\begin{center} \includegraphics[clip,scale=0.25]{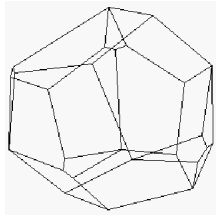}
\vspace{-.5cm}
\end{center}
\caption{Fundamental region for Weeks manifold. }

\label{varw}
\end{figure}

This manifold was discovered independently by Weeks \cite{weeks}
and Matveev-Fomenko \cite{MtF}, and is the manifold with the smallest
volume (in units of $a^{3}$) known, $V=0.942707...a^{3}$. Its fundamental
region is an $18$-face polyhedron, shown in Figure \ref{varw}. The
radius of the inscribed sphere is $R_{inradius}=0.519162...a$.

\begin{figure}[H]
\begin{center} \includegraphics[clip,scale=0.35]{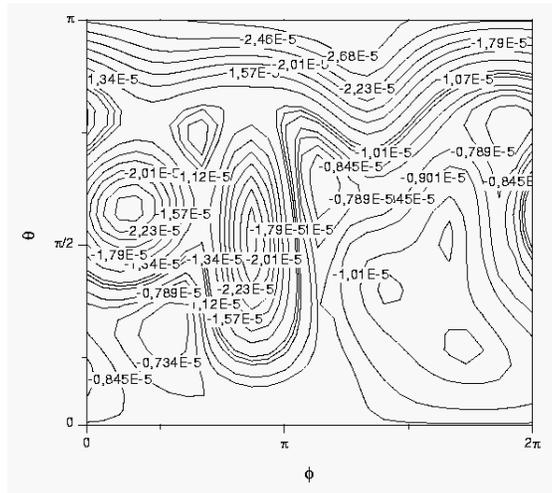}
\end{center}
\caption{$\rho_{C}$ for Weeks universe. }

\label{2}
\end{figure}

The vacuum expectation value of the $00$-component of the energy-momentum
tensor, $\rho_{C}=T_{\mu\nu}u^{\mu}u^{\nu}$, as seen by a comoving
observer, is shown in Figure \ref{2}.
%
\newpage
\subsection{Thurston Manifold}

\begin{figure}[H]
\begin{center} \includegraphics[clip,scale=0.3]{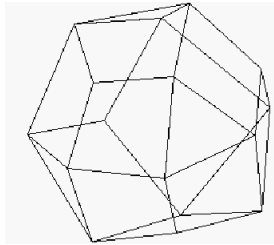}
\end{center}
\caption{Fundamental region for Thurston manifold.}

\label{rfvth}
\end{figure}

It was discovered by the field medalist William Thurston \cite{Trst}.
This manifold possesses a fundamental region of $16$ faces, its volume
is $V=0.981369...a^{3}$ (Figure \ref{rfvth}), and $R_{inradius}=0.535437...a$.

\begin{figure}[H]
\begin{center} \includegraphics[clip,scale=0.35]{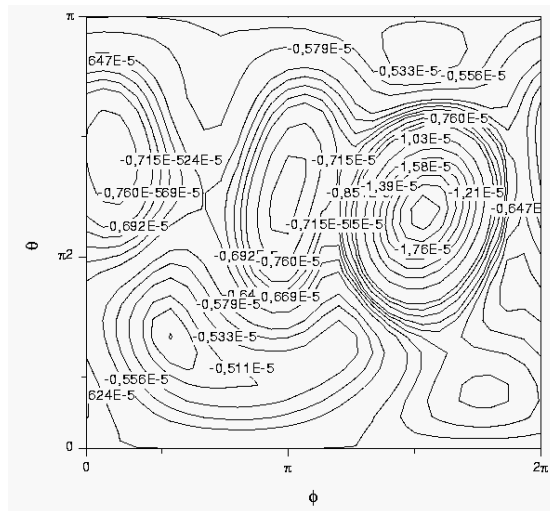}
\end{center}
\caption{$\rho_{C}$ for Thurston universe.}

\label{3}
\end{figure}

Figure \ref{3} shows the value of $\rho_{C}=T_{\mu\nu}u^{\mu}u^{\nu}=T_{00},$
as seen by a comoving observer. %
\newpage
\subsection{Best Manifold}

\begin{figure}[H]
\begin{center} \includegraphics[clip,scale=0.3]{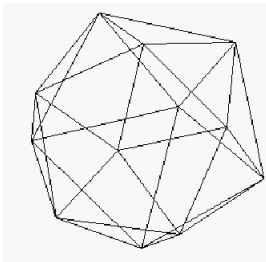}
\end{center}
\caption{Fundamental region for Best manifold.}

\label{rfvb}
\end{figure}

This manifold was discovered as a by-product of a study of finite
subgroups of $SO(1,3)$ by a geometrical aproach \cite{best}. Its
fundamental region is an icosahedron with $V=4.686034...a^{3}$ and
$R_{inradius}=0.868298...a$, shown in Figure \ref{rfvb}.

\begin{figure}[H]
\begin{center} \includegraphics[clip,scale=0.35]{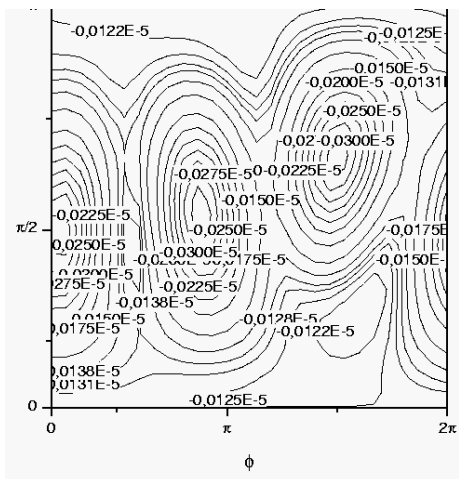}
\end{center}
\caption{$\rho_{C}$ for Best universe.}
\label{4}
\end{figure}

The vacuum expectation value of $\,\,\rho_{C}=T_{\mu\nu}u^{\mu}u^{\nu}$,
as seen by a comoving observer, is shown in Figure \ref{4}.


\subsection{Seifert-Weber Manifold \label{D}}

\begin{figure}[H]
\begin{center} \includegraphics[clip,scale=0.3]{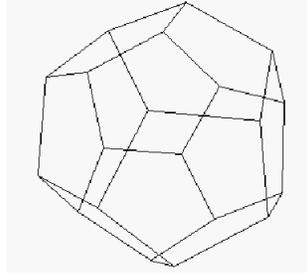}
\end{center}
\caption{Fundamental region for Seifert-Weber manifold.}
\label{rfvsw}
\end{figure}

For this manifold, which was discovered by Weber and Seifert \cite{WS},
$V=11.199065...a^{3}$, $R_{inradius}=0.996384...a,$ and the fundamental
region is a dodecahedron (Figure \ref{rfvsw}).

\begin{figure}[H]
\begin{center} \includegraphics[clip,scale=0.35]{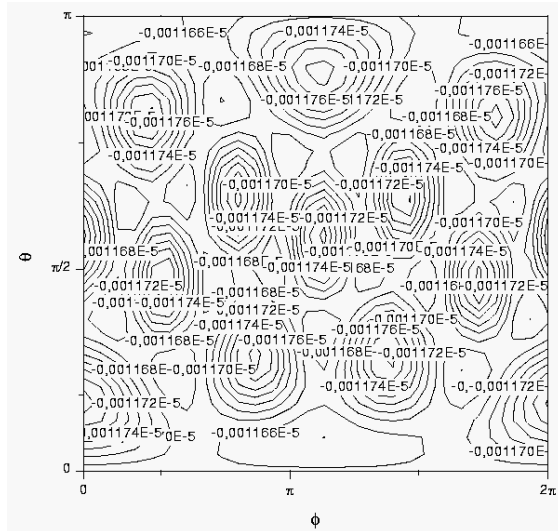}
\end{center}
\caption{$\rho_{C}$ for Seifert-Weber universe.}
\label{5}
\end{figure}

Figure \ref{5} shows the value of $\rho_{C}=T_{\mu\nu}u^{\mu}u^{\nu},$
as seen by a comoving observer.


\section{Conclusions}

We explicitly evaluated the distribution of the vacuum energy density
of a conformally coupled massive scalar field, for static universes
with compact spatial sections of negative curvature and increasing
volume: Weeks, Thurston, Best, and Seifert-Weber manifolds. As a specific
example, we chose $m=0.4$ for the mass of the scalar field, and $a=10$
for the radius of curvature. The values of the Casimir energy density
$\rho_{C}$ on a sphere of proper (geodesic) radius $d=3.90035..$
inside the fundamental polyhedron for each of these manifolds are
shown in Figures \ref{2}, \ref{3}, \ref{4}, and \ref{5}. In all
these cases it can be seen that there is a spontaneous generation
of low multipolar components. As expected, the effect becomes weaker
for increasing volume universes.

\section*{Acknowledgments}

D. M. would like to thank the brazilian agencies Finatec and Funpe
for partial support. This was presented as an invited talk, for which I would 
like to thank the MG10 organizing committee, in particular M. J. Rebouças and 
M. Demianski.

\end{document}